\documentclass[twocolumn,linenumbers,trackchanges]{aastex631}
\usepackage{units}
\usepackage{amsmath}
\usepackage{multirow}
\usepackage{graphicx}

\newcommand{\messierdist}{\textcolor{black}{\unit[6.7]{Mpc}}}
\newcommand{\messierdistsigma}{\textcolor{black}{\unit[0.9]{Mpc}}}

\newcommand{\OSWstart}{\textcolor{black}{2023-05-13T19:49:35}}
\newcommand{\OSWend}{\textcolor{black}{2023-05-18T19:49:35}}
\newcommand{\firstdetmjd}{\textcolor{black}{60082.82611}}
\newcommand{\firstdetcv}{\textcolor{black}{18.76}}
\newcommand{\firstdetcverr}{\textcolor{black}{0.25}}
\newcommand{\lastnondetmjd}{\textcolor{black}{60082.66041667}}
\newcommand{\lastnondetclear}{\textcolor{black}{20.4}}
\newcommand{\oswduration}{\textcolor{black}{five}}
\newcommand{\coincduration}{\textcolor{black}{\unit[0.8]{days}}}
\newcommand{\analysisduration}{\textcolor{black}{\unit[0.68]{days}}}
\newcommand{\oswcoverage}{\textcolor{black}{\unit[14]{\%}}}
\newcommand{\percentveto}{\textcolor{black}{15 \%}}

\newcommand{\faploudest}{\textcolor{black}{0.75}}

\newcommand{\ifartenyr}{\textcolor{black}{$1$\, event in $10$\, years}}
\newcommand{\fapifartenyr}{\textcolor{black}{$1.9 \times 10^{-4}$}}
\newcommand{\ellipninty}{\textcolor{black}{1.08}}
\newcommand{\egwninty}{\textcolor{black}{\unit[$1 \times 10^{-4}$]{$\mathrm{M_\odot c^2}$}}}
\newcommand{\lumninty}{\textcolor{black}{\unit[$2.6 \times 10^{-4}$]{$\mathrm{M_\odot c^2/s}$}}}
\newcommand{\distnintyneut}{\textcolor{black}{\unit[6.9]{kpc}}}

\newcommand{\distnintyPow}{\textcolor{black}{\unit[29.9]{kpc}}}

\newcommand{\egwixfejjnew}{\textcolor{black}{49}}
\newcommand{\powerixfejjnew}{\textcolor{black}{36}}
\newcommand{\ellixfejjnew}{\textcolor{black}{6.8}}
\newcommand{\ellipnintyhigh}{\textcolor{black}{$3.6 \times 10^2$}}
\nolinenumbers

\begin{document}

\title{Search for gravitational waves emitted from SN\,2023ixf}



\author[0000-0003-4786-2698]{A.~G.~Abac}
\affiliation{Max Planck Institute for Gravitational Physics (Albert Einstein Institute), D-14476 Potsdam, Germany}
\author{R.~Abbott}
\affiliation{LIGO Laboratory, California Institute of Technology, Pasadena, CA 91125, USA}
\author{I.~Abouelfettouh}
\affiliation{LIGO Hanford Observatory, Richland, WA 99352, USA}
\author{F.~Acernese}
\affiliation{Dipartimento di Farmacia, Universit\`a di Salerno, I-84084 Fisciano, Salerno, Italy}
\affiliation{INFN, Sezione di Napoli, I-80126 Napoli, Italy}
\author[0000-0002-8648-0767]{K.~Ackley}
\affiliation{University of Warwick, Coventry CV4 7AL, United Kingdom}
\author{S.~Adhicary}
\affiliation{The Pennsylvania State University, University Park, PA 16802, USA}
\author[0000-0002-4559-8427]{N.~Adhikari}
\affiliation{University of Wisconsin-Milwaukee, Milwaukee, WI 53201, USA}
\author[0000-0002-5731-5076]{R.~X.~Adhikari}
\affiliation{LIGO Laboratory, California Institute of Technology, Pasadena, CA 91125, USA}
\author{V.~K.~Adkins}
\affiliation{Louisiana State University, Baton Rouge, LA 70803, USA}
\author[0000-0002-8735-5554]{D.~Agarwal}
\affiliation{Universit\'e catholique de Louvain, B-1348 Louvain-la-Neuve, Belgium}
\affiliation{Inter-University Centre for Astronomy and Astrophysics, Pune 411007, India}
% ...

\collaboration{3000}{The LIGO Scientific Collaboration, the Virgo Collaboration, and the KAGRA Collaboration}



\begin{abstract}
We present the results of a search for gravitational-wave transients associated with core-collapse supernova SN\,2023ixf, 
which was observed in the galaxy Messier 101 via optical emission on 2023 May 19th, during the LIGO-Virgo-KAGRA 15th Engineering Run.
We define a five-day on-source window during which an accompanying gravitational-wave signal may have occurred. No gravitational waves have been identified in data when at least two gravitational-wave observatories were operating, which covered $\sim$ \oswcoverage\, of this \oswduration-day window. We report the search detection efficiency for various possible gravitational-wave emission models. Considering the distance to M101 (\messierdist), we derive constraints on the gravitational-wave emission mechanism of core-collapse supernovae across a broad frequency spectrum, ranging from \unit[50]{Hz} to \unit[2]{kHz} where we assume the gravitational-wave emission occurred when coincident data are available in the on-source window. Considering an ellipsoid model for a rotating proto-neutron star, our search is sensitive to gravitational-wave energy \egwninty\, and luminosity \lumninty\, for a source emitting at \unit[82]{Hz}. These constraints are around an order of magnitude more stringent than those obtained so far with gravitational-wave data. The constraint on the ellipticity of the proto-neutron star that is formed is as low as \ellipninty, at frequencies above 1200 Hz, surpassing past results.

\end{abstract}


\section{Introduction}
\label{sec:intro}

The direct detection of gravitational waves (GWs) from a binary black hole merger \citep{LIGOScientific:2016aoc} started the field of GW astronomy, and was followed by similar mergers \citep{LIGOScientific:2018mvr,LIGOScientific:2020ibl,LIGOScientific:2021usb,KAGRA:2021vkt}. Two years later, the merger of two neutron stars was observed both with GWs and across the electromagnetic spectrum \citep{LIGOScientific:2017vwq,LIGOScientific:2017ync}, leading to the birth of GW multi-messenger astronomy. More recently, the observation of mergers of mixed systems \citep{LIGOScientific:2021qlt, LIGOScientific:2024elc} is allowing measurement of the merger rates of all types of compact binary systems \citep{KAGRA:2021duu}.

Core-collapse supernovae (CCSNe) are the explosions of massive stars -- masses above 8 $\mathrm{M_{\odot}}$\, at the end of their evolution -- leading to the production of neutron stars and black holes~\citep{Burrows:1995ww,Kotake:2005zn,Janka:2012wk}.
CCSNe are astrophysical sources with  multi-messenger emission, having historically been observed over the electromagnetic spectrum and, for SN 1987A, also with low-energy neutrinos \citep{Kamiokande-II:1987idp,Bionta:1987qt,1987ESOC...26..237A}.
However, the GW emission of CCSNe is still undetected.
The combination of GW and neutrino observations can provide information about the collapse and the onset of the explosion, since both messengers are emitted from the core very soon after the collapse and have negligible interactions with the surrounding matter~\citep{Janka:2012wk}.
On the other hand, electromagnetic emission is produced in the outer layers of the star and is delayed.

The GW emission from CCSNe is weaker than the emission from compact binary mergers, making it detectable by the advanced generation of detectors only for nearby supernovae \citep{Gossan:2015xda,Szczepanczyk:2021bka,KAGRA:2021tnv}.
The most likely opportunity for observations are Galactic CCSNe, but the expected rate is of the order of one or two per century \citep{Bergh:1991ek,Cappellaro:1993ns,Tammann:1994ev,Diehl:2006cf,Li:2010kc,Adams:2013ana}.
However, due to the large uncertainties of the progenitors and GW emission models, we carry out searches for GW emission from CCSNe out to distances of \unit[20]{Mpc} \citep{LIGOScientific:2016jvu,LIGOScientific:2019ryq,Szczepanczyk:2023ihe}. 

SN\,2023ixf was identified in Messier 101 (M101) during its rise, making it one of the closest type II CCSNe observed. The two LIGO observatories were in observing mode during the fifteenth Engineering Run (ER15) of the LIGO-Virgo-KAGRA network \citep{LIGOScientific:2014pky,VIRGO:2014yos,KAGRA:2020tym}.
In this article, we report the results of the search for GWs and the new constraints on GW emission obtained with SN\,2023ixf.

\section{SN\,2023ixf and ER15 data}

\subsection{Summary of SN\,2023ixf multi-messenger observations} 
\label{subsec:multi}

SN\,2023ixf (RA = 14:03:38.562, DEC = +54:18:41.94, J2000) was discovered on 2023 May 19 by \citet{2023TNSTR1158....1I} with a clear (unfiltered) magnitude of 14.9 in the host galaxy M101 (NGC 5457, Pinwheel Galaxy).
M101 is at a distance of about \messierdist~(see Sec.~\ref{subsec:dist}), making SN\,2023ixf one of the nearest CCSNe observed in recent years. In addition, this galaxy is a well-observed object with an extensive set of pre-discovery observations.
SN\,2023ixf was quickly classified as a type II supernova a few hours after the discovery \citep{2023TNSAN.119....1P}.
Due to the prompt discovery and the close distance, SN\,2023ixf was the target of extensive electromagnetic coverage.
The optical light curve shows a rise to a maximum at about five days, followed by a plateau lasting for about one month, and a slow decline later \citep{Hiramatsu:2023inb,Hosseinzadeh:2023ixa,Li:2023vux,Sgro:2023zko,Teja:2023hcm,Yamanaka:2023gbr}.
The early spectroscopic observations show flash ionization features of hydrogen, helium, nitrogen, carbon and a temperature increase not explained by pure shock cooling, suggesting a delayed shock breakout in a dense circumstellar medium \citep{Berger:2023jcl, Bersten:2023vct,Bostroem:2023dvn,Grefenstette:2023dka,Chandra:2023kda,Guetta:2023mls,Hiramatsu:2023inb,Hosseinzadeh:2023ixa,2023RNAAS...7..169K,Jacobson-Galan:2023ohh,Kilpatrick:2023pse,Li:2023vux,Martinez:2023cmx,Murase:2023chr,Niu:2023tmz,Pledger:2023ick,2023arXiv230910022Q,Smith:2023kqo,Teja:2023hcm,VanDyk:2023tns,Vasylyev:2023mtl,Xiang:2023xos,Yamanaka:2023gbr,Zimmerman:2023mls}.
The earliest detections of X-ray and radio emission occurred four days \citep{Grefenstette:2023dka} and one month \citep{2023ATel16091....1M} after the discovery, respectively.
The hard X-ray \citep{Grefenstette:2023dka} and soft X-ray \citep{Chandra:2023kda, panjkov_24} observations suggest a high and decreasing neutral hydrogen column density close to SN\,2023ixf.
SN\,2023ixf was not detected in gamma-rays \citep{2023ATel16075....1M} or in neutrinos \citep{2023ATel16043....1T,2023ATel16070....1N,IceCube:2023wku}.

\subsection{Nature and mass of progenitor}
\label{subsec:prog}

A large set of M101 pre-discovery imaging observations from ground-based telescopes, Hubble Space Telescope and Spitzer Space Telescope suggest the nature of the SN\,2023ixf progenitor to be a dusty and variable red supergiant, with an estimated mass ranging from 8 to 20 $\mathrm{M_{\odot}} $ \citep{Dong:2023wsc,Flinner:2023jph,Hiramatsu:2023inb,Jencson:2023bxz,Neustadt:2023fao,Niu:2023tmz,Pledger:2023ick,Ransome:2023kps,Soraisam:2023ktz,VanDyk:2023tns,Xiang:2023xos,Ferrari:2024avb,Moriya:2024pzs}.

The circumstellar medium could have been produced by an enhancement in the mass loss before the SN explosion, but several archival investigations did not find any pre-explosion outburst in the years before the discovery \citep{Dong:2023wsc,Flinner:2023jph,Jencson:2023bxz,Neustadt:2023fao,Ransome:2023kps,Soraisam:2023ktz}, while detecting amplitude pulsations \citep{Kilpatrick:2023pse,Soraisam:2023ktz}.

\subsection{M101 distance}
\label{subsec:dist}

The distance of the supernova host galaxy is relevant to constrain the GW energy emission.
Since astronomical distances are estimated using a broad range of methods, we have considered the available published values to estimate the distance to M101.
More precisely, we have considered the distance estimations reported in the NASA Extragalacic Database \citep{1991ASSL..171...89H}, a total number of 115 measurements using 12 different methods: Cepheids, Planetary Nebulae Luminosity Function, Supernova Ia, Tip of Red Giant Branch, SN II optical, Brightest Stars, Tully-Fisher relation, M Stars, RSV Stars, S Dor stars, H II region diameter and SN II radio. 
We adopt the median to the remaining 115 data points, \messierdist, with a standard deviation of \messierdistsigma, as the estimated distance of SN\,2023ixf.

\begin{figure*}[t]
  \centering
  \includegraphics[width=\textwidth]{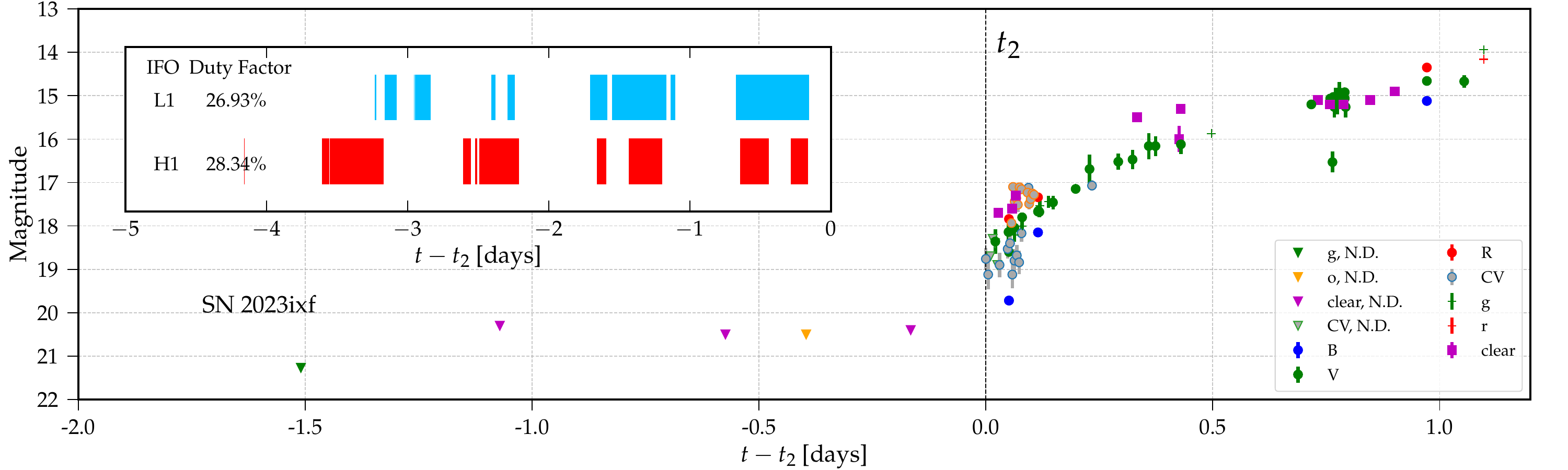}
  \caption{Early evolution of SN\,2023ixf covering different photometric bands (B, V, R, g, o) and unfiltered observations (CV, clear); N.D. marks non detections; inset: LIGO Hanford (H1) and Livingston (L1) detectors duty cycle within the OSW described in the text. Photometric data sources: Transient Name Server Astronotes, Astronomical Telegrams, AAVSO,  \citet{Sgro:2023zko,Li:2023vux}.
  }.
 \label{fig:osw}
\end{figure*}

\subsection{On-source window}
\label{subsec:osw}

The on-source window is the time interval containing the core bounce and the following GW emission. We denote the start and end times of this interval as $t_1$ and $t_2$, respectively. Due to the availability of well-sampled public photometric data of SN\,2023ixf, the on-source window could be estimated using the early photometric observations that include the non-detections before the rise to peak brightness as shown in Fig. \ref{fig:osw}. The first detection is MJD = \firstdetmjd, at a CV magnitude of \firstdetcv$\pm$\firstdetcverr ~\citep{2023TNSAN.150....1C}, following the last pre-discovery observation at MJD = \lastnondetmjd, clear magnitude $>$ \lastnondetclear ~\citep{2023TNSAN.130....1M}.
For SN\,2023ixf, $t_2$ is well approximated by the first detection, while $t_1$ involves the delay between collapse and shock breakout, whose time falls between $t_2$ and the latest pre-discovery observation.
The time delay depends on many properties of the progenitor, including its mass. Considering the large spread in mass estimations and the relation between mass and time delay found by \cite{Barker:2021iyr} (Fig. 6), we have adopted a conservative maximal on-source window duration of \oswduration~days, from \OSWstart\, to \OSWend\, UTC.

\subsection{ER15 data}
ER15 took place at the LIGO Livingston and Hanford Observatories from 2023 April 7 to 2023 May 24 following a period of upgrades and commissioning that improved the detectors' sensitivity from the previous observing run. During ER15, the observatories collect data as if it were a normal observing run with the exception that calibration, commissioning, and noise investigations are performed. These studies are concentrated near the beginning of ER15 and taper to an as-needed basis towards the last week. The collapse of SN\,2023ixf likely happened during this end period of ER15 as did the time period spanned by this search. 

The uncertainty in the strain calibration has been found to be similar to previous observing runs~\citep{Sun2021}. Its effect on the search is marginal and thus ignored.
Within the on-source window, the two LIGO observatories were operating jointly for $\sim$ \coincduration.

Transient data artifacts, referred to as glitches, contaminate the data and can affect the confidence estimation of candidate events. The search has been carried out with strain channels {\tt L1:GDS-CALIB\_STRAIN\_CLEAN\_AR} and {\tt H1:GDS-CALIB\_STRAIN\_CLEAN\_AR} where {\tt CLEAN} means some of the well identified noise sources have been removed~\citep{KAGRA:2023pio}. Data quality studies reveal auxiliary channels that are insensitive to GWs and have a strong correlation to the glitches in the output of the detector. These times of poor data quality are then removed (vetoed) \citep{Davis2021}, representing \percentveto\, of the coincident time within the on-source window. This gives the analysis time of $\sim$ \analysisduration.

\section{Search} 
\label{sec:method}

\begin{figure*}[hbt] 
  \begin{minipage}[c][][t]{0.495\textwidth}
    \vspace*{\fill}
    \flushleft
    \includegraphics[width=0.96\linewidth]{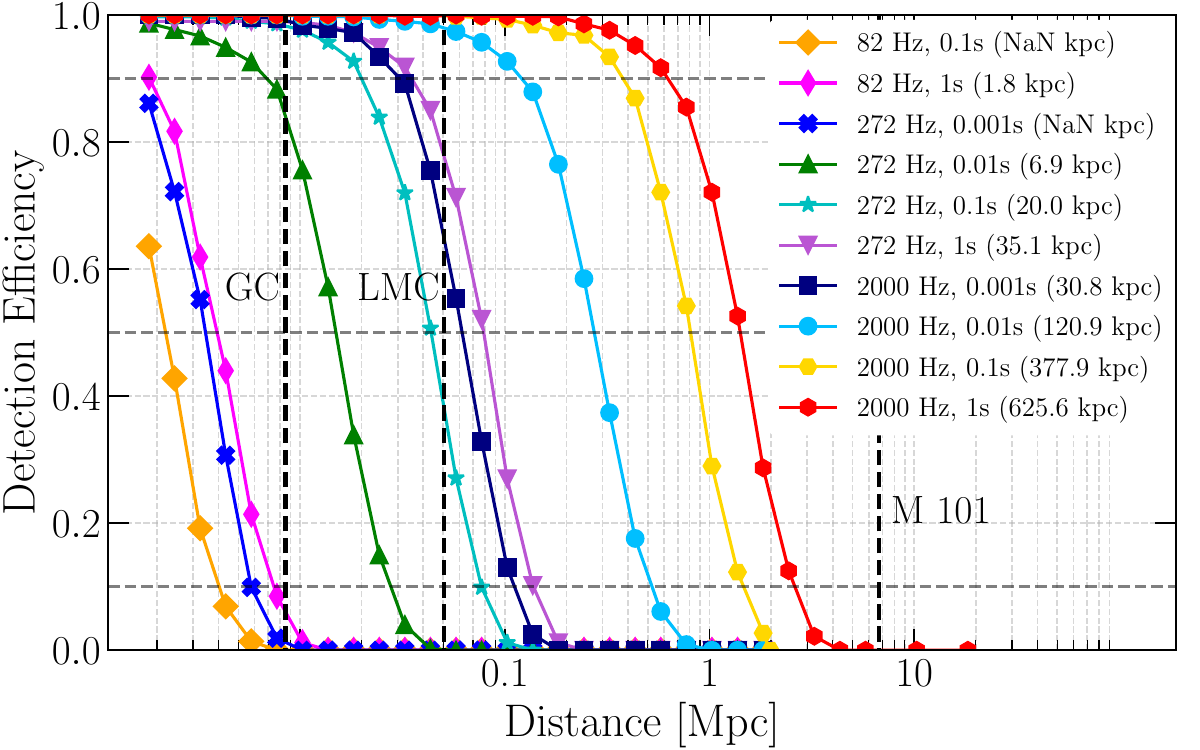}
  \end{minipage}%
  \begin{minipage}[c][][t]{0.495\textwidth}
    \vspace*{\fill}
    \flushright
    \includegraphics[width=0.96\linewidth]{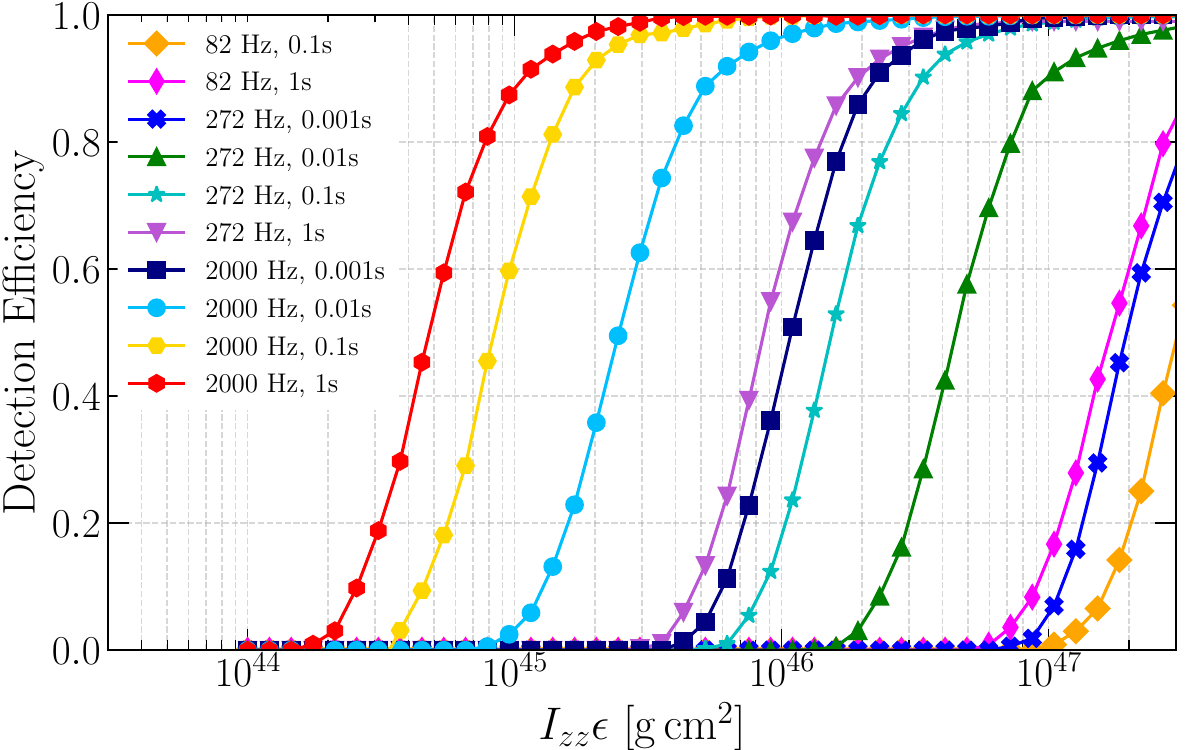}
  \end{minipage}
    \caption {Detection efficiency of a GW source in the direction of M101, and within the on-source window of SN\,2023ixf, for a FAR of less than \ifartenyr. The long-lasting bar-mode GW emission model is considered for various peak frequencies and durations. On the left panel $I_{zz}\epsilon$ is fixed to \unit[$0.1 \times 10^{45}$]{$\mathrm{g\,cm^2}$} and the detection efficiency is shown as function of the distance to the source. 
    Horizontal dashed lines show 10\%, 50\%, and 90\% detection efficiencies and the Galactic Center (GC), Large Magellanic Cloud (LMC) and M101 distances are shown as references. On the right panel, the distance is fixed to M101 and the detection efficiency is shown as function of $I_{zz}\epsilon$.
   }
    \label{fig:effcurves}
\end{figure*} 

\subsection{Coherent WaveBurst}

We use coherent WaveBurst (cWB), a model-agnostic search algorithm, for the detection and reconstruction of transient GW signals~\citep{cwb}. The algorithm identifies GW transients by searching for excess power in spectrograms and reconstructs coherent signals in multiple detectors. In previous CCSN searches~\citep{Szczepanczyk:2022urr}, spectrograms were obtained with the Wilson-Daubechies-Meyer wavelet transform~\citep{Necula:2012zz}. The SN\,2023ixf analysis uses the high-resolution wavescan transform~\citep{Klimenko:2022nji} that utilizes both the excess-power and cross-power statistics for the identification of GW signals and enables more accurate reconstruction of the signal waveforms~\cite{Mishra:2024zzs}. The signal detection statistic $\eta_{\rm 0}$ is defined as $\eta_{\rm 0}=\sqrt{E_c}$ where $E_c$ is the total coherent energy across the detector network \citep{cwb}. To further separate GW signals from the noise, the triggers are then re-ranked with a reduced statistic $\eta_\mathrm{r} = \eta_\mathrm{0}\cdot W_{\mathrm{XGB}}$, where $W_{\mathrm{XGB}}$ is the XGBoost classification penalty factor which ranges between 0 (noise-like) and 1 (signal-like)~\citep{Szczepanczyk:2022urr,Mishra:2021tmu}. For this search, the XGBoost algorithm uses the SN\,2023ixf sky location to improve detection sensitivity. 

\subsection{CCSN models} 
\label{sec:waveforms}

To test the sensitivity of the search, we use a range of different waveforms from numerical simulations, that span the expected progenitor parameter space. For non-rotating sources we use the \unit[15]{$\mathrm{M_{\odot}}$} SFHx s15 model from \citet{kuroda2016} [Kur+16 s15], the \unit[15]{$\mathrm{M_{\odot}}$} D15 model from \cite{mezzacappa_23} [Mez+23 D15], the \unit[20]{$\mathrm{M_{\odot}}$} mesa20\_pert model from \cite{oconnor2018} [Oco+18 m20p], the \unit[18]{$\mathrm{M_{\odot}}$} s18 model from \cite{powell2019} [Pow+19 s18], the \unit[40]{$\mathrm{M_{\odot}}$} NR model from \cite{pan2021} [Pan+21 NR], and the \unit[25]{$\mathrm{M_{\odot}}$} s25 model from \cite{radice2019} [Rad+19 s25]. For examples of progenitors at the lower mass end, we include model he3.5 from \cite{powell2019} [Pow+19 he3.5], which is an ultra-stripped progenitor with a \unit[3.5]{$\mathrm{M_{\odot}}$} helium core, and the \unit[13]{$\mathrm{M_{\odot}}$} s13 model from \cite{radice2019} [Rad+19 s13]. 

We also include waveforms from more energetic types of explosions. We include the \unit[50]{$\mathrm{M_{\odot}}$} s50 model from \cite{Kuroda:2021eiv} [Kur+22 s50], as an example of a CCSN explosion powered by a first-order quantum-chromodynamics phase transition. We include several rotating models, as the rotation can significantly increase the GW amplitude. They are the \unit[40]{$\mathrm{M_\odot}$} model SR from \cite{pan2021} [Pan+21 SR], the \unit[15]{$\mathrm{M_{\odot}}$} s15fr model from \cite{Andresen_19} [And+19 s15fr], and the \unit[39]{$\mathrm{M_{\odot}}$} helium star model m39 from \cite{Powell:2020cpg} [Pow+20 m39]. We also include a few models that include both rapid rotation and magnetic fields, as this can result in powerful magnetorotational explosions. They are the \unit[39]{$\mathrm{M_{\odot}}$} m39\_B12 model from \cite{Powell_23} [Pow+23 B12], and model 3d\_signal\_O from \cite{Obergaulinger:2019iyi} [Obe+20 signal\_O].

We also consider a phenomenological emission model related to the development of long-lasting bar-mode instabilities inside the proto-neutron star (PNS) \citep{ott:10dcc, Gossan:2015xda}. Assuming the PNS is well modelled as a triaxial ellipsoid rotating around the $z$\, axis, one can approximate the GW emission with sine-Gaussian waveforms

\begin{equation}
\label{eqn:wvf}
\begin{split}
h_{+}(t) &= \frac{1}{2} ~h_0~ [1+\cos^2 \iota]~{\rm e}^{\frac{-t^2}{\tau^2}} ~\cos(2\pi f_0 t)~, \\
h_{\times}(t) &= h_0 ~\cos \iota ~{\rm e}^{\frac{-t^2}{\tau^2}} ~\sin(2\pi f_0 t)~,\\
\end{split}
\end{equation}
where 
\begin{equation}
    h_0 = \frac{2}{D} \frac{G}{c^4}\frac{I_{zz} \epsilon}{2}~ (2\pi f_0)^2~,
\end{equation}
$I_{zz}$ and $\epsilon$ are the moment of inertia and ellipticity of the ellipsoid, $f_0$ is twice the rotation frequency, $D$ is the source distance and $\iota$ is the inclination angle of the $z$ axis with respect to the line of sight. 
$I_{zz} \epsilon$\, is a free parameter. Throughout the paper, we consider the canonical value for neutron stars $I_{zz}=$ \unit[$10^{45}$]{$\mathrm{g\,cm^2}$}~\citep{Paschalidis:2016vmz} and keep $\epsilon$ as a free parameter.

\section{Results} \label{sec:search} 
\subsection{Search result and background estimation}
\label{sec:bkg}

The detector data contains a variety of transient noise sources that contribute to the search background. 
To assess the significance of each trigger, we compute the false-alarm rate (FAR), which estimates the frequency of noise triggers mistakenly identified as potential GW events. Within the on-source window, the trigger with the lowest FAR is considered a GW event candidate. In this search, the lowest FAR event candidate has a FAR of 2.11 per day, giving a false-alarm probability of $1 - \mathrm{e}^{-T_{\rm obs} \times \mathrm{FAR}} = $ \faploudest; i.e., a probability of \faploudest\, that noise alone would produce a trigger of this FAR or lower ($T_{\rm obs}=\analysisduration$). This suggests that this trigger is likely due to noise.

\subsection{Detection efficiency}
\label{sec:eff}

To evaluate the search sensitivity, we take the signal models described in Sec.~\ref{sec:waveforms} and randomize the source orientation such that it is uniformly distributed over a sphere. Then we add waveforms to the detector coincident data within the on-source window for the sky location of SN\,2023ixf.
We compute the search detection efficiency, defined as the fraction of detected signals with FAR lower than \ifartenyr. This FAR corresponds to a false alarm probability of \fapifartenyr. At the distance of SN\,2023ixf none of the 14 models from numerical simulations are detected. In Table \ref{tab:eff_dist} we report the distance at which we recover 90\% of the added signals for all 14 CCSN models.
The distances reach up to \distnintyneut\, for the non-rotating explosions, which means that signals from non-rotating or slow-rotating progenitor CCSN could be missed within the Galaxy for sources further than the Galactic Center.
The distances for the more extreme models are around a factor 4 larger than for the non-rotating explosions, exceeding the Galaxy boundaries -- \distnintyPow\, for Pow+23 B12 model -- but without reaching the large Magellanic Cloud distance. Finally, for the explosion driven by a first-order quantum-chromodynamics phase transition, a detection of 90\%  is never achieved because only one polarization is extracted from the 2D numerical simulations.


\begin{table}[htbp]
\centering
\caption{Distance of the 90\% detection efficiency reached with CCSN waveform models for a FAR of \ifartenyr. Values in bold represent the farthest distance reached for each family of models. For the 2D Kur+22 s50 model, detection efficiency remains lower than 90\% whatever the distance because there is only one polarization. We report the 50\% detection efficiency instead that is marked with $^*$.}

\begin{tabular}{l|l|l}
\hline
\multicolumn{2}{c|}{Waveform Models} &
  Distance [kpc] \\ \hline
\multirow{8}{*}{\parbox{.2\columnwidth}{Non-rotating models}}  &  Kur+16 s15 &  \textbf{6.9} \\
 & Mez+23 D15       & 2.9           \\
 & Oco+18 m20p      & 1.0           \\
 & Pow+19 s18       & 5.5           \\
 & Pow+19 he3.5     & 2.8           \\
 & Rad+19 s13       & 0.6           \\
 & Rad+19 s25       & 5.8         \\
 & Pan+21 NR        & 6.6         \\ \hline
\multirow{5}{*} {\parbox{.2\columnwidth}{Rotating models}} & And+19 s15fr &  1.8 \\
& Obe+20 Signal\_O & 13.4          \\ 
& Pan+21 SR        & 18.2          \\ 
& Pow+20 m39       & 19.6         \\ 
& Pow+23 B12       & \textbf{29.9} \\ \hline
Phase & \multirow{3}{*}{Kur+22 s50} &  \multirow{3}{*}{\textbf{$8.9^*$}} \\
transition & & \\
model & &\\\hline
\end{tabular}

\label{tab:eff_dist}

\end{table}

Fig.~\ref{fig:effcurves} shows the detection efficiency for long-lasting bar-mode waveforms with frequencies between \unit[82]{Hz} and \unit[2]{kHz} and signal durations between \unit[1]{ms} and \unit[1]{s}. The right plot is as a function of $I_{zz} \epsilon$ for a source at the location of SN\,2023ixf. The left plot is as a function of distance for $I_{zz} \epsilon \sim$\, \unit[$0.1\times 10^{45}$]{$\mathrm{g\,cm^2}$}. The sensitivity increases with the signals peak frequency and duration. For instance, we could detect at 90\%\, confidence level a  signal lasting \unit[1]{s} at \unit[2]{kHz} for $I_{zz} \epsilon \sim$\, \unit[$10^{45}$]{$\mathrm{g\,cm^2}$}. For lower frequencies, if we assume the canonical value $I_{zz} \sim$\, \unit[$10^{45}$]{$\mathrm{g\,cm^2}$}, the source would need to be highly deformed ($\epsilon \gg 1$).

\section{Constraints}
\label{sec:constraints}

Assuming the GW emission occurred when coincident data are available in the on-source window, we establish constraints on several quantities characterizing a core collapse, including emitted GW energy, luminosity, and PNS ellipticity, considering the long-lasting bar-mode model. These bar-mode instabilities are sometimes present in simulations at low rotational kinetic energy over gravitational potential energy ratio ($T/|W|$).  

\begin{figure}[hbt] 
    \centering
    \includegraphics[width=0.95\columnwidth]{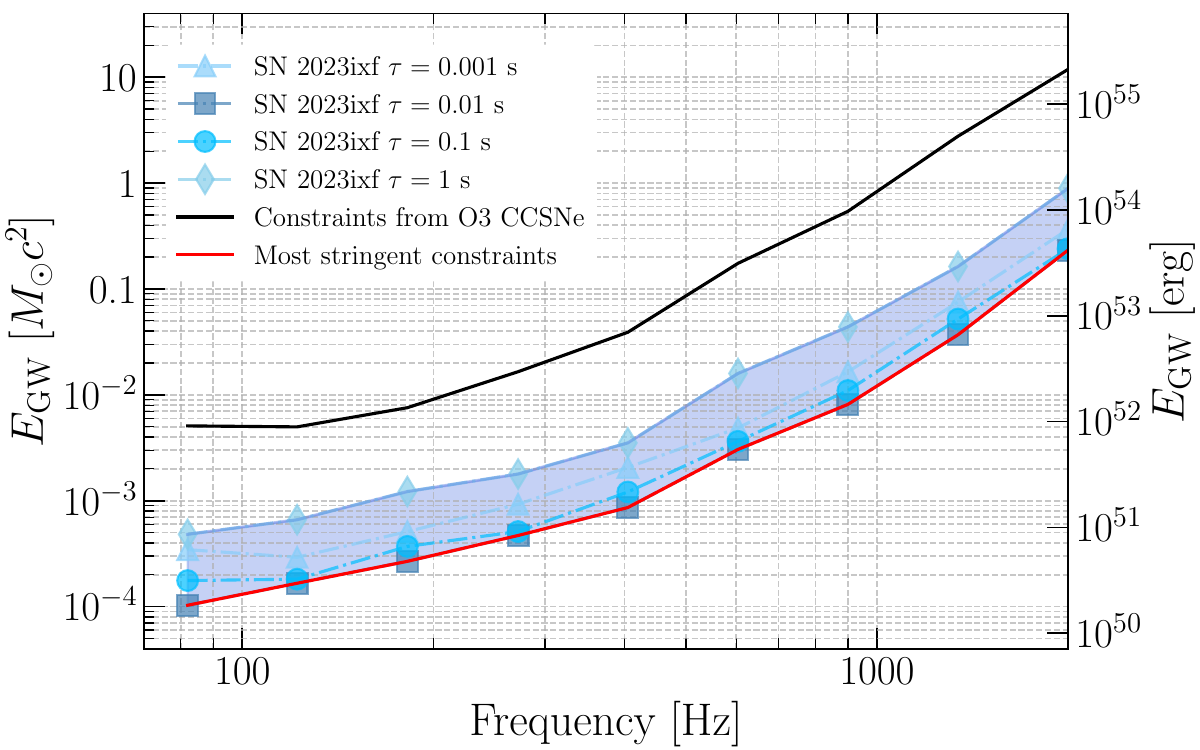}
    \includegraphics[width=0.95\columnwidth]{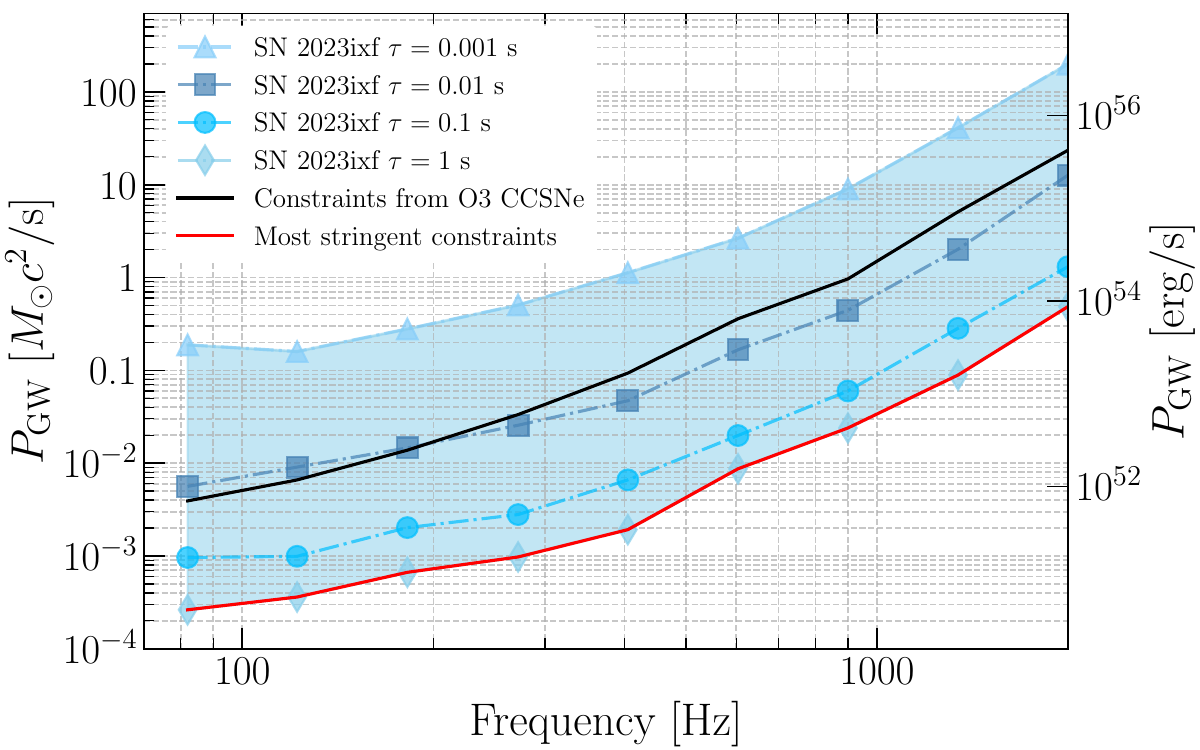}
    \caption{     
  GW energy ($E_\mathrm{GW}$) and luminosity ($P_\mathrm{GW}$) as a function of the frequency for bar-mode signals with a detection efficiency of 90\% and a FAR of \ifartenyr. The shaded region contains combined results from all analyzed models for SN\,2023ixf. }
\label{fig:eng} 
\end{figure}

\subsection{Constraints on GW energy and luminosity}
\label{sec:egw}

Assuming a rotating core, the emitted GW energy is~\citep{Sutton:2013ooa}

\begin{equation}
\begin{split}
  E_\mathrm{GW} &= \frac{2}{5} \frac{\pi^2 c^3}{G}D^2 f^2_0 \int_{-\infty}^{\infty} [h_+^2(t)+h_{\times}^2(t)]~dt \\
  &= \frac{2}{5} \frac{\pi^2 c^3}{G}  \sqrt{\frac{\pi}{2}} \tau D^2 f^2_0 h^2_0
\end{split}
  \label{eqn:energy}
\end{equation}
where the GW strain squared integral is computed for an optimally oriented source. 

The GW luminosity is the ratio between the emitted GW energy and the duration of the emission. We define the duration as the time interval $\tau_{90}$ that contains 90\% of the energy such that the GW average luminosity is given by

\begin{equation}
  P_\mathrm{GW} = \frac{0.9 ~E_\mathrm{GW}}{\tau_{90}}\,.
\end{equation}
For the sine-Gaussians of Eq. (\ref{eqn:wvf}) $\tau_{90}=1.65~\tau$.
Considering the $h_0$ value corresponding to 90\% detection efficiency we derive constraints on $E_\mathrm{GW}$ and $P_\mathrm{GW}$ shown in Fig.~\ref{fig:eng}. The shaded region contains results from all long-lasting bar-mode models. At \unit[82]{Hz} the more stringent energy constraints are $\sim$ \egwninty. Fig.~\ref{fig:eng} also shows the constraints derived from GW searches targetting CCSNe during the third observing run (O3) of the LIGO-Virgo-KAGRA network.
The constraints with SN\,2023ixf are $\sim$ \egwixfejjnew\, times more stringent than for O3 CCSNe over the whole frequency range. 
For the emitted GW average luminosity shown in the bottom panel, the constraints are \lumninty\, for signals at \unit[82]{Hz} and \unit[1]{s} long. They are a factor of $\sim$ \powerixfejjnew\, more stringent compared to the O3 CCSNe over the whole frequency range.
This upper limit is between 2 and 5 orders of magnitude larger than the GW average luminosity predicted by numerical simulations for the different explosion mechanisms \citep{Szczepanczyk:2021bka} showing that we could soon exclude some of the most optimistically luminous models with a closer CCSN. It is also $\sim 7$\, orders of magnitude larger than the bolometric luminosity reported in \cite{Zimmerman:2023mls}.

\subsection{Constraints on PNS ellipticity}
\label{sec:ell}

As shown in Sec.~\ref{sec:waveforms}, the amplitude of the GW signal emitted by a rotating PNS can be parametrized by its ellipticity and its moment of inertia given by the relation 

\begin{equation}
    I_{zz} \epsilon = \frac{Dc^4}{G (2 \pi f_0)^2} h_0
\end{equation}

\begin{figure}[!hbt]
  \centering
  \includegraphics[width=0.95\columnwidth]{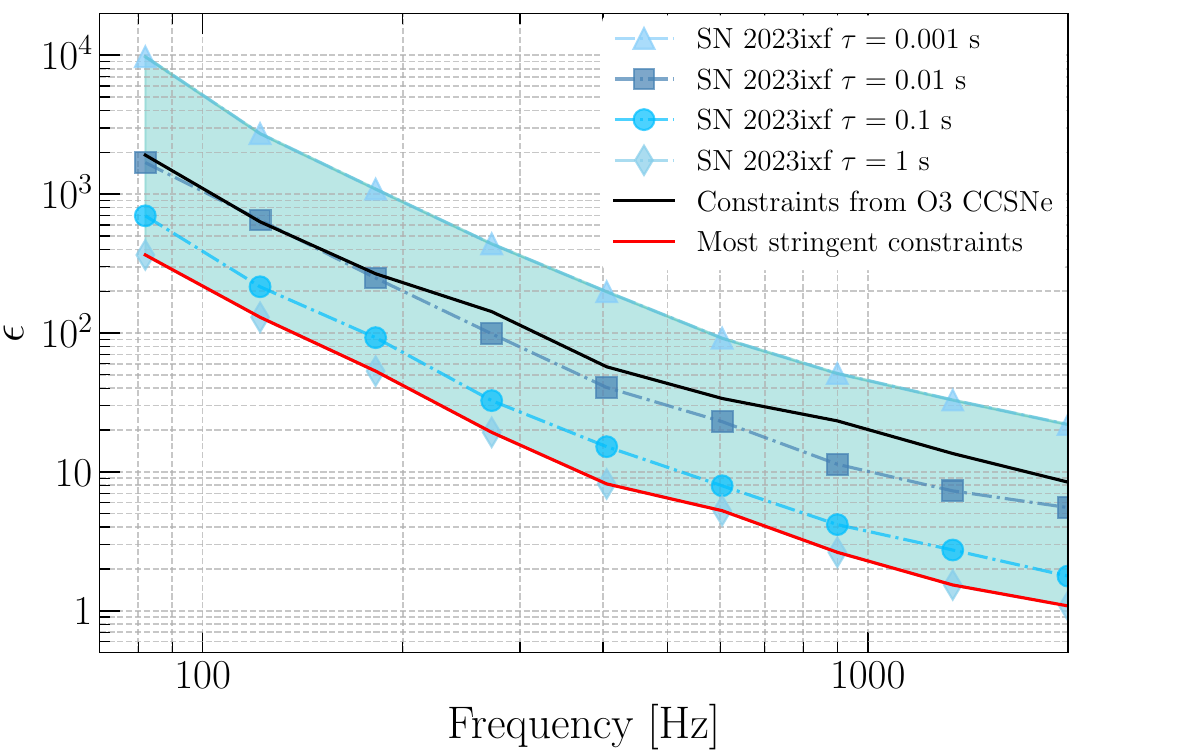}
  \caption{PNS ellipticity as a function of the frequency for bar-mode signals with a detection efficiency of 90\% and a FAR of 1 per 10 years. The moment of inertia $I_{zz}$\, is fixed to \unit[$10^{45}$]{$\mathrm{g\,cm}^2$}. The shaded region contains combined results from all analyzed bar-mode models for SN\,2023ixf. }
\label{fig:ell}
\end{figure}

Fig.~\ref{fig:ell} reports the ellipticity for a range of bar-mode GW signal frequencies and durations for a detection efficiency of 90\%. 
The most stringent constraints on ellipticity are obtained for the signals with $\tau=1$\,s, ranging from \ellipnintyhigh\, at the lowest search frequency to \ellipninty\, at 2\,kHz. The $\epsilon$ constraints get stricter with longer signals. 
Over the whole frequency range, the constraints given by SN\,2023ixf on the ellipticity are $\sim$ \ellixfejjnew\, more stringent than for O3 CCSNe.

\section{Summary and Discussion} 
\label{sec:sum}

We present the results of a search for GW signals coincident with SN\,2023ixf, which was observed during the LIGO-Virgo-KAGRA Engineering Run 15, 2023 April 24 to 2023 May 24.
No significant GW candidates were identified within the $\sim$ \oswcoverage\, of the on-source window where coincident good quality GW data are available. 
With different CCSN waveform models, we quantify the search sensitivity by estimating the distances at which 90\% of the GW simulated signals are detected.
The reported distances are up to \distnintyneut\, for non-rotating explosions, and up to \distnintyPow\, for rapidly rotating models. These distance sensitivities have been obtained using the FAR of \ifartenyr\,. 
We derive constraints on the GW energy, luminosity, and PNS ellipticity, which are the most stringent that GW detector data have achieved to date. Assuming the PNS is well modelled as a rotating triaxial ellipsoid whose moment of inertia along the rotation axis is fixed to $I_{zz}$=\unit[$10^{45}$]{$\mathrm{g\,cm^2}$}, we find that the ellipticity should be lower than \ellipninty. This value, obtained for an hypothetical \unit[1]{s}-long signal at \unit[2]{kHz}, is one order of magnitude larger than plausible estimates ($\sim 0.1$) derived from simulations where bar-mode instabilities are present \citep{Shibagaki:2020ksk,Obergaulinger:2021,Bugli:2022mlq}.

Despite the large distance of SN\,2023ixf, this event probes regions of the bar-mode instabilities parameter space that are physically interesting. On the other hand, in the case of a neutrino-driven, magnetorotational or more exotic explosion model such as first-order quantum-chromodynamics phase transition, we show that for detecting GWs from CCSNe, events within the Local group are still the best prospect.
\\


\newif\ifcoreonly
\newif\ifkagra
\newif\ifheader
\coreonlyfalse \kagratrue  \headerfalse
%
\ifheader
\begin{center}{\bf\Large
\ifkagra
Conference proceedings acknowledgements for \\ the LIGO Scientific Collaboration, the Virgo Collaboration and the KAGRA Collaboration
\else
Conference proceedings acknowledgements for \\ the LIGO Scientific Collaboration and the Virgo Collaboration
\fi
}\end{center}
\fi
This material is based upon work supported by NSF’s LIGO Laboratory which is a major facility
fully funded by the National Science Foundation.
The authors also gratefully acknowledge the support of
the Science and Technology Facilities Council (STFC) of the
United Kingdom, the Max-Planck-Society (MPS), and the State of
Niedersachsen/Germany for support of the construction of Advanced LIGO 
and construction and operation of the GEO\,600 detector. 
Additional support for Advanced LIGO was provided by the Australian Research Council.
The authors gratefully acknowledge the Italian Istituto Nazionale di Fisica Nucleare (INFN),  
the French Centre National de la Recherche Scientifique (CNRS) and
the Netherlands Organization for Scientific Research (NWO), 
for the construction and operation of the Virgo detector
and the creation and support  of the EGO consortium. 
The authors also gratefully acknowledge research support from these agencies as well as by 
the Council of Scientific and Industrial Research of India, 
the Department of Science and Technology, India,
the Science \& Engineering Research Board (SERB), India,
the Ministry of Human Resource Development, India,
the Spanish Agencia Estatal de Investigaci\'on (AEI),
the Spanish Ministerio de Ciencia, Innovaci\'on y Universidades,
the European Union NextGenerationEU/PRTR (PRTR-C17.I1),
the ICSC - CentroNazionale di Ricerca in High Performance Computing, Big Data
and Quantum Computing, funded by the European Union NextGenerationEU,
the Comunitat Auton\`oma de les Illes Balears through the Direcci\'o General de Recerca, Innovaci\'o i Transformaci\'o Digital with funds from the Tourist Stay Tax Law ITS 2017-006,
the Conselleria d'Economia, Hisenda i Innovaci\'o, the FEDER Operational Program 2021-2027 of the Balearic Islands,
the Conselleria d'Innovaci\'o, Universitats, Ci\`encia i Societat Digital de la Generalitat Valenciana and
the CERCA Programme Generalitat de Catalunya, Spain,
the Polish National Agency for Academic Exchange,
the National Science Centre of Poland and the European Union – European Regional
Development Fund;
the Foundation for Polish Science (FNP),
the Polish Ministry of Science and Higher Education,
the Swiss National Science Foundation (SNSF),
the Russian Science Foundation,
the European Commission,
the European Social Funds (ESF),
the European Regional Development Funds (ERDF),
the Royal Society, 
the Scottish Funding Council, 
the Scottish Universities Physics Alliance, 
the Hungarian Scientific Research Fund (OTKA),
the French Lyon Institute of Origins (LIO),
the Belgian Fonds de la Recherche Scientifique (FRS-FNRS), 
Actions de Recherche Concertées (ARC) and
Fonds Wetenschappelijk Onderzoek – Vlaanderen (FWO), Belgium,
the Paris \^{I}le-de-France Region, 
the National Research, Development and Innovation Office Hungary (NKFIH), 
the National Research Foundation of Korea,
the Natural Science and Engineering Research Council Canada,
Canadian Foundation for Innovation (CFI),
the Brazilian Ministry of Science, Technology, and Innovations,
the International Center for Theoretical Physics South American Institute for Fundamental Research (ICTP-SAIFR), 
the Research Grants Council of Hong Kong,
the National Natural Science Foundation of China (NSFC),
the Leverhulme Trust, 
the Research Corporation,
the National Science and Technology Council (NSTC), Taiwan,
the United States Department of Energy,
and
the Kavli Foundation.
The authors gratefully acknowledge the support of the NSF, STFC, INFN and CNRS for provision of computational resources.

\ifcoreonly\else
\fi
\ifkagra
This work was supported by MEXT,
the JSPS Leading-edge Research Infrastructure Program,
JSPS Grant-in-Aid for Specially Promoted Research 26000005,
JSPS Grant-in-Aid for Scientific Research on Innovative Areas 2905: JP17H06358,
JP17H06361 and JP17H06364,
JSPS Core-to-Core Program A,
Advanced Research Networks,
JSPS Grants-in-Aid for Scientific Research (S) 17H06133 and 20H05639,
JSPS Grant-in-Aid for Transformative Research Areas (A) 20A203: JP20H05854,
the joint research program of the Institute for Cosmic Ray Research,
the University of Tokyo,
the National Research Foundation (NRF),
the Computing Infrastructure Project of Global Science experimental Data hub
Center (GSDC) at KISTI,
the Korea Astronomy and Space Science Institute (KASI),
the Ministry of Science and ICT (MSIT) in Korea,
Academia Sinica (AS),
the AS Grid Center (ASGC) and the National Science and Technology Council (NSTC)
in Taiwan under grants including the Rising Star Program and Science Vanguard
Research Program,
the Advanced Technology Center (ATC) of NAOJ,
and the Mechanical Engineering Center of KEK.
\fi


\ifcoreonly\else
\fi

\ifcoreonly\else
\fi

\ifcoreonly\else
\fi
\ifcoreonly\else
{\it We would like to thank all of the essential workers who put their health at risk during the COVID-19 pandemic, without whom we would not have been able to complete this work.}
\fi

\bibliography{snsearch}{}
\bibliographystyle{aasjournal}

\end{document}